\documentclass{aa}
\usepackage{txfonts}
\usepackage{graphicx}
\usepackage{subcaption}
\usepackage[colorlinks=true,allcolors=blue]{hyperref}

\begin{document} 

\titlerunning{}
\authorrunning{Mountrichas et al.}
\titlerunning{SF of AGN and non AGN as a function of environment and morphology}

\title{The relation of cosmic environment and morphology with the star formation and stellar populations of AGN and non-AGN galaxies}

\author{G. Mountrichas\inst{1}, G. Yang\inst{2,3}, V. Buat\inst{4,5}, B. Darvish\inst{6}, M. Boquien\inst{7}, Q. Ni\inst{8}, D. Burgarella\inst{4} and L. Ciesla\inst{4}}
          
     \institute {Instituto de Fisica de Cantabria (CSIC-Universidad de Cantabria), Avenida de los Castros, 39005 Santander, Spain
              \email{gmountrichas@gmail.com}
              \and
               Kapteyn Astronomical Institute, University of Groningen, P.O. Box 800, 9700 AV Groningen, The Netherlands
               \and
               SRON Netherlands Institute for Space Research, Postbus 800, 9700 AV Groningen, The Netherlands
                  \and
             Aix Marseille Univ, CNRS, CNES, LAM Marseille, France. 
              \and
                 Institut Universitaire de France (IUF)
                  \and
                     University of California, Riverside, 900 University Avenue, Riverside, CA 92521, USA
                     \and
                     Centro de Astronom\'ia (CITEVA), Universidad de Antofagasta, Avenida Angamos 601, Antofagasta, Chile   
                     \and
                      Max-Planck-Institut f\"{u}r extraterrestrische Physik (MPE), Gie{\ss}enbachstra{\ss}e 1, D-85748 Garching bei M\"unchen, Germany}

\abstract{In this work, we study the relation of cosmic environment and morphology with the star-formation (SF) and the stellar population of galaxies. Most importantly, we examine if this relation differs for systems with active and non-active supermassive black holes. For that purpose, we use 551 X-ray detected active galactic nuclei (AGN) and 16,917 non-AGN galaxies in the COSMOS-Legacy survey, for which the surface-density field measurements are available. The sources lie at redshift of $\rm 0.3<z<1.2$, probe X-ray luminosities of $\rm 42<log\,[L_{X,2-10keV}(erg\,s^{-1})]<44$ and have stellar masses, $\rm 10.5<log\,[M_*(M_\odot)]<11.5$. Our results show that isolated AGN (field) have lower SFR compared to non AGN, at all L$_X$ spanned by our sample. However, in denser environments (filaments, clusters), moderate L$_X$ AGN ($\rm log\,[L_{X,2-10keV}(erg\,s^{-1})]>43$) and non-AGN galaxies have similar SFR. We, also, examine the stellar populations and the morphology of the sources in different cosmic fields. For the same morphological type, non-AGN galaxies tend to have older stellar populations and are less likely to have undergone a recent burst in denser environments compared to their field counterparts. The differences in the stellar populations with the density field are, mainly, driven by quiescent systems. Moreover, low L$_X$ AGN present negligible variations of their stellar populations, in all cosmic environments, whereas moderate L$_X$ AGN have, on average, younger stellar populations and are more likely to have undergone a recent burst, in high density fields. Finally, in the case of non-AGN galaxies, the fraction of bulge-dominated (BD) systems increases with the density field, while BD AGN are scarce in denser environments. Our results are consistent with a scenario in which a common mechanism, such as mergers, triggers both the SF and the AGN activity.}

   
\maketitle

\section{Introduction}

It is widely accepted that galaxies and the supermassive black hole (SMBHs) in their centres co-evolve. This symbiosis has been demonstrated in a variety of ways. For instance, tight correlations have been found between the black hole mass and various galaxy properties \citep[e.g., bulge or stellar mass, stellar velocity dispersion;][]{Magorrian1998, Ferrarese2000, Tremaine2002, Kormendy2013}. Moreover, both the SMBH activity and star formation peak at similar cosmic times \citep[$\rm z\sim 2$; e.g.,][]{Boyle2000, Sobral2013}. In addition, both the black hole and galaxy growth are fed by cold gas.

A popular method to examine the co-evolution of galaxies and SMBHs is to study the relation between the star-formation rate (SFR) of the galaxy and the accreting power of the SMBH in their centre. For the latter, the X-ray luminosity, L$_X$, is used as a proxy in galaxies that host active SMBHs. These galaxies are known as active galactic nuclei (AGN). Although, this has been the topic of many studies \citep[e.g.][]{Lutz2010, Page2012, Lanzuisi2017}, more information can be gained regarding the effect of the AGN feedback on the host galaxy, by comparing the SFR of AGN with that of non-AGN systems, with similar stellar mass, M$_*$ and redshift \citep[e.g.][]{Rosario2013, Mullaney2015, Masoura2018, Bernhard2019, Florez2020, Masoura2021, Koutoulidis2022, Pouliasis2022}. Recent studies that applied this method found that for galaxies with $\rm 10.5<log\,[M_*(M_\odot)]<11.5$ the SFR of low-to-moderate L$_X$ ($\rm 42<log\,[L_{X,2-10keV}(erg\,s^{-1})]<44$) AGN tends to be lower, or at most equal to the SFR of non-AGN, star-forming galaxies, while an enhancement of $\sim 30\%$ was found for the SFR of galaxies that host higher L$_X$ AGN compared to non-AGN systems. Nevertheless, in both lower mass and more massive galaxies the SFR of AGN was similar to that of non-AGN systems \citep{Mountrichas2021c, Mountrichas2022a, Mountrichas2022b}. These results were also supported by studying the stellar populations of low-to-moderate L$_X$ AGN compared to non-AGN systems \citep{Mountrichas2022c}. 

A limitation of the above works is that they do not take into account the cosmic environment of the two populations (i.e., AGN and non-AGN systems). A significant number of studies, though, have investigated if and how the cosmic web affects the star-formation of galaxies (non-AGN), both in the local Universe and at high redshifts \citep[e.g.][]{Brambila2023, Song2023}. At low redshifts ($\rm z<1$), most studies suggest that quiescent (Q) galaxies preferentially live in dense environments, such as groups and clusters, whereas star-forming galaxies are more commonly found in the field \citep[e.g.][]{Lin2014, Erfanianfar2016}. A plausible scenario to interpret these results is that cold gas is removed in satellite galaxies due to their interaction with the warm intra-cluster medium. At high redshifts, though, this picture seems to be either reversed \citep[e.g.,][]{Elbaz2007, Santos2014} or no difference has been found in the SFR of galaxies that live in different density fields \citep[e.g.,][]{Scoville2013, Cooke2023}. Furthermore, a number of studies have shown that one of the galaxy properties that are mostly influenced by the cosmic environment is the galaxy morphology \citep[e.g.,][]{Dressler1980, Fasano2015}. Nevertheless, these studies do not take into account the activity of the SMBH. In other words, they do not differentiate between AGN and non-AGN galaxies.

In this work, we compare the SFR and star-formation histories (SFH) of AGN and non-AGN systems, for different cosmic environments (field, filaments, clusters) and morphologies (bulge-dominated, BD, and non bulge-dominated, non-BD). Our goal is to examine the effect of AGN feedback on the host's star-formation in different density fields and compare the role of the cosmic web and morphology on the stellar population of AGN and non-AGN galaxies. For that purpose, we use sources detected in the COSMOS field and lie at $\rm z\sim 1$. We compute M$_*$ by applying spectral energy distribution (SED) fitting using the CIGALE code, probing a range of $\rm 10.5<log\,[M_*(M_\odot)]<11.5$. First, we compare the SFR of galaxies that host AGN with their non-AGN counterparts as a function of the cosmic environment. Then, we compare the effect of the density field and morphology on the stellar populations of AGN and non-AGN galaxies. Finally, we discuss our results and describe our main conclusions. Throughout this work, we assume a flat $\Lambda$CDM cosmology with $H_ 0=70.4$\,Km\,s$^{-1}$\,Mpc$^{-1}$ and $\Omega _ M=0.272$ \citep{Komatsu2011}.

\section{Data}
\label{sec_data}

\subsection{The X-ray dataset}

In this study, we use sources that lie within the COSMOS field \citep{Scoville2007}. The X-ray AGN have been observed by the COSMOS-Legacy survey \citep{Civano2016}. This is a 4.6\,Ms {\it{Chandra}} program that covers 2.2\,deg$^2$ of the COSMOS region. The central area has been observed with an exposure time of $\approx 160$\,ks while the remaining area has an exposure time of $\approx 80$\,ks. The limiting depths are $2.2 \times 10^{-16}$, $1.5 \times 10^{-15}$ , and $8.9 \times 10^{-16}\,\rm erg\,cm^{-2}\,s^{-1}$ in the soft (0.5-2\,keV), hard (2-10\,keV), and full (0.5-10\,keV) bands, respectively. The X-ray catalogue includes 4016 sources. \cite{Marchesi2016} matched the X-ray sources with optical and infrared counterparts using the likelihood ratio technique \citep{Sutherland_and_Saunders1992}. Of the sources, 97\%  have an optical and IR counterpart and a photometric redshift (photo-{\it z}) and $\approx 54\%$  have spectroscopic redshift (spec-{\it z}). Photo-{\it z} have been derived from high-quality ultraviolet-to-infrared data \citep[up to 32 bands;][]{Laigle2016}. Hardness ratios ($\rm HR=\frac{H-S}{H+S}$, where H and S are the net counts of the sources in the hard and soft band, respectively) were estimated for all X-ray sources using the Bayesian estimation of hardness ratios method \citep[BEHR;][]{Park2006}. The intrinsic column density, N$\rm _H$, for each source was then calculated using its redshift and assuming an X-ray spectral power law with slope $\Gamma=1.8$. This information is available in the catalogue presented in \cite{Marchesi2016}. In this work, we use sources within the UltraVISTA region \citep[1.38\,deg$^2$][]{McCracken2012} of the COSMOS field. There are 1718 X-ray sources within this region with $\rm log\,[L_{X,2-10keV}(erg\,s^{-1})]>42$.

%The photometric redshifts available in their catalogue, have been produced following the procedure described in \cite{Salvato2011}. Different libraries of templates have been used depending on the X-ray flux of the source and the morphological and photometric properties of the associated counterpart. The best fit was derived using the LePhare code \citep{Arnouts1999, Ilbert2006}. The accuracy of photometric redshifts is found at $\sigma_{\Delta z/(1+z_{spec})}=0.03$. The fraction of outliers ($\Delta z/(1+z_{zspec})>0.15$) is $\approx 8\%$. 

In our analysis, we use the X-ray dataset used in \cite{Mountrichas2022a}. Specifically, we use the same strict photometric selection criteria with them, to make sure that only sources with the best photometric coverage are included in our analysis ($u, g, r, i, z, J, H, K_s$, IRAC1, IRAC2, and MIPS/24, where IRAC1, IRAC2, and MIPS/24 are the [3.6]\,$\mu$m, [4.5]\,$\mu$m, and 24 $\mu$m photometric bands of Spitzer). We also apply the same reliability requirements with those described in Sect. 3.2 in \cite{Mountrichas2022a} to make certain that the analysis is restricted to those sources with the most reliable host galaxy properties (see Sect. \ref{sec_analysis}). We, also, use only sources that meet the mass completeness limits, as described in detail in sections 2.1, 3.2 and 3.4 in \cite{Mountrichas2022a}. Specifically, for the redshift range spanned by the sample, the stellar mass completeness limits are $\rm log\,[M_{*,95\%lim}(M_\odot)]= 9.13$ and 9.44 at $\rm 0.5<z<1.0$ and $\rm 1.0<z<1.5$, respectively. There are 1161 X-ray sources that meet all the above selection criteria \citep[Table 2 in][]{Mountrichas2022a}. 

One goal of this study is to examine the SFR of galaxies that host AGN in different cosmic environments. \cite{Yang2018b}, used sources in the COSMOS field, at $\rm 0.3<z<3.0$ and studied the dependence of the black hole growth on the cosmic environment. For that purpose, they adopted the `weighted adaptive kernel smoothing' method to construct the surface-density field that probes sub-Mpc physical scales \citep{Darvish2015}. The method was applied to all sources in the COSMOS field, that is for both non-AGN galaxies and X-ray systems. The end product of this analysis is the calculation of the dimensionless overdensity parameter ($1+\delta=\frac{\Sigma}{\Sigma _{median}}$, where $\Sigma$ is the surface number density, in units of Mpc$^{-2}$, and $\Sigma_{median}$ is the median $\Sigma$ at each redshift). The method is described in detail in Sect. 2.3.1 in \cite{Yang2018b} \citep[see also][]{Darvish2015}. Based on the density field estimates, the cosmic-web is, then, extracted using the multiscale environment filter (MMF) algorithm \citep[e.g.,][]{AragonCalvo2007, Darvish2014, Darvish2017}. The main idea is to measure the geometry of the density field around each point at each redshift. If the geometry is similar to that of a typical cluster or filament, then the point's environment is classified as cluster or filament. Otherwise, the point is classified as the field \citep[for more details, see Sect. 2.3.2 in][]{Yang2018b}.

To add to our X-ray sources the information about the field density and cosmic environment, we cross-match the 1161 sources with the catalogue used in \cite{Yang2018b}. This results in 1005 X-ray AGN. The missing AGN (1161 vs. 1005) are due to the fact that in \cite{Yang2018b}, they filter out sources near ($<1$\.Mpc) the edge of the field, since density measurements for these sources are unreliable \citep[see also][]{Darvish2017}.

\subsection{The galaxy control sample}

The main goal of this work, is to compare the SFR and SFH of AGN host galaxies in different cosmic environments with the SFR of non-AGN systems, in similar density fields. Towards this end, we use the galaxy control sample presented in \cite{Mountrichas2022a} (see their Sect. 2.2). We apply the same selection criteria used for the X-ray dataset and the same mass completeness limits, which result in 89375 sources \citep[Table 2 in][]{Mountrichas2022a}. We note that X-ray sources have been excluded from the galaxy control sample, as well as sources with a strong AGN component, calculated by the SED fitting \citep[see next section and Sect. 3.3 in][]{Mountrichas2022a}. To include information about the cosmic environment of the sources in the galaxy control sample, we cross-match it with the catalogue presented in \cite{Yang2018b}. There are 76,251 common sources between the two datasets.

\section{Galaxy properties}
\label{sec_analysis}

In this section, we describe how we obtain information about the properties of the sources used in our analysis. Specifically, we present how we measure the SFR and M$_*$ of AGN and non-AGN galaxies and how we retrieve knowledge on their stellar populations and morphology.

\subsection{Calculation of SFR and M$_*$}

The (host) galaxy properties of both the X-ray AGN and the galaxies in the control sample have been calculated via SED fitting, using the CIGALE code \citep{Boquien2019, Yang2020, Yang2022}. The SED fitting analysis is described in detail in Section 3.1 in \cite{Mountrichas2022a}. In brief, the galaxy component is modelled using a delayed SFH model with a function form $\rm SFR\propto t \times exp(-t/\tau)$. A star formation burst is included \citep{Ciesla2017, Malek2018, Buat2019} as a constant ongoing period of star formation of 50\,Myr. Stellar emission is modelled using the single stellar population templates of \cite{Bruzual_Charlot2003} and is attenuated following the \cite{Charlot_Fall_2000} attenuation law. To model the nebular emission, CIGALE adopts the nebular templates based on \cite{Inoue2011}. The emission of the dust heated by stars is modelled based on \cite{Dale2014}, without any AGN contribution. The AGN emission is included using the SKIRTOR models of \cite{Stalevski2012, Stalevski2016}. CIGALE has the ability to model the X-ray emission of galaxies. In the SED fitting process, the observed L$_X$ in the $2-10$\,keV band are used, provided by the \cite{Marchesi2016}. The parameter space used is shown in Table 1 in \cite{Mountrichas2022a}. The reliability of the SFR measurements, both in the case of AGN and non-AGN systems, has been examined in detail in our previous works and, in particular, in Sect. 3.2.2 in \cite{Mountrichas2022a}.

\begin{figure}
\centering
  \includegraphics[width=0.9\linewidth, height=7cm]{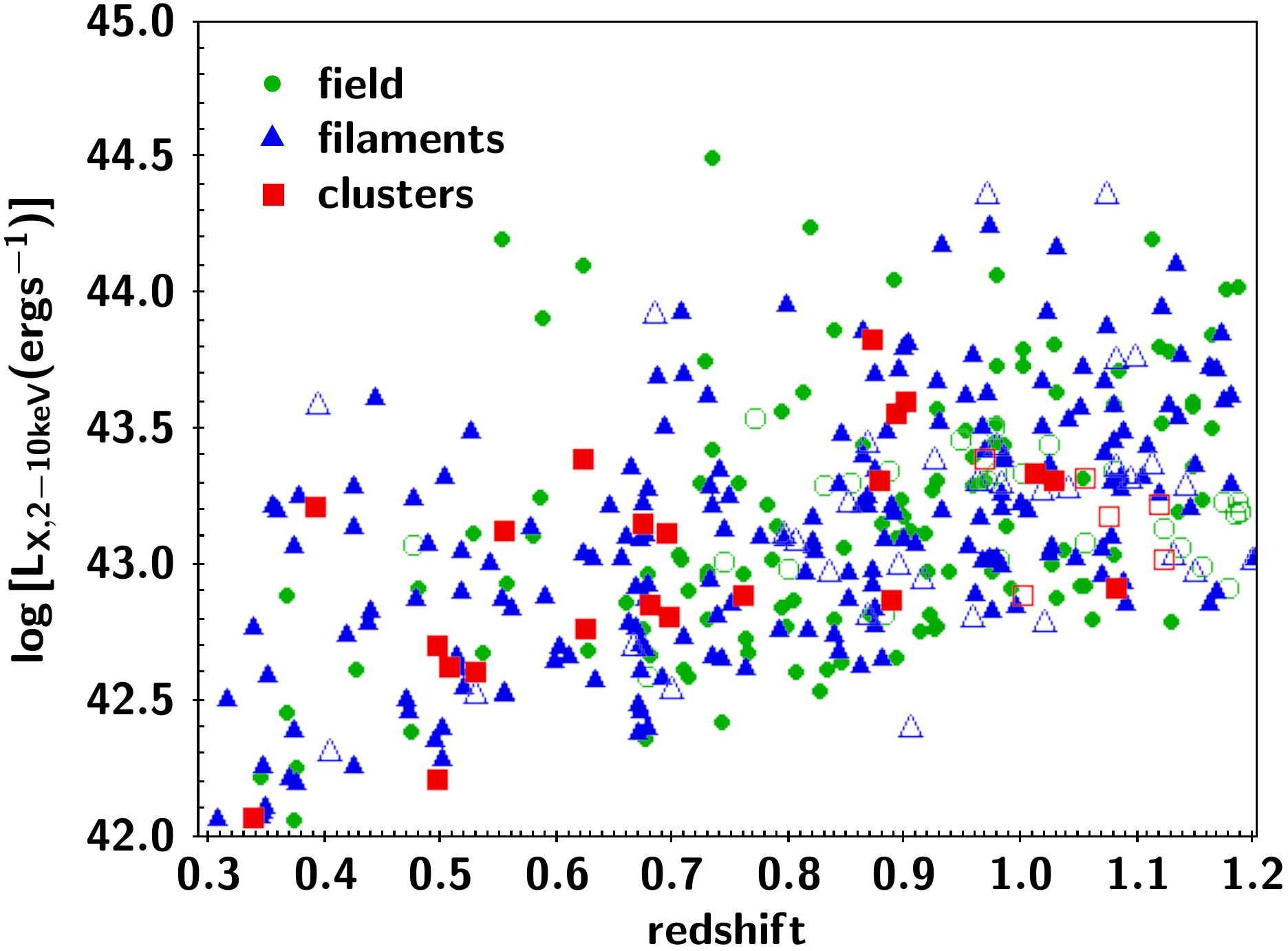}
  \caption{The distribution of the 551 X-ray AGN in the L$_X-$redshift plane. Different colours and symbols correspond to sources associated with different cosmic environments, as indicated in the legend. Open symbols present AGN with photo-{\it{z}}.}
  \label{fig_lx_redz}
\end{figure}

In \cite{Mountrichas2021c, Mountrichas2022a, Mountrichas2022b}, we found that the SFR$_{norm}$-L$_X$ relation depends on the stellar mass range probed by the sources. Specifically a flat SFR$_{norm}$-L$_X$ relation was found for the least and most massive systems ($\rm log\,[M_*(M_\odot)]<10.5$ and $\rm log\,[M_*(M_\odot)]>11.5$), with SFR$_{norm}\sim 1$. Albeit, for intermediate stellar masses ($\rm 10.5<log\,[M_*(M_\odot)]<11.5$) SFR$_{norm}$ was found to be SFR$_{norm}\leq 1$ at low-to-moderate L$_X$ ($\rm log\,[L_{X,2-10keV}(erg\,s^{-1})]<44$) whereas at higher L$_X$, SFR$_{norm}>1$  \citep[e.g., see Fig. 5 in][]{Mountrichas2022b}. Among the 1005 X-ray AGN in our dataset, $\sim 80\%$ (809 sources) have $\rm 10.5<log\,[M_*(M_\odot)]<11.5$. Therefore, there are not enough sources to probe lower or higher M$_*$ in a statistically robust manner. Thus, we restrict our analysis to AGN (and galaxies) with $\rm 10.5<log\,[M_*(M_\odot)]<11.5$. Furthermore, sources in our datasets (AGN and non-AGN) associated with filaments or the field, span redshifts up to 2.5. Nevertheless, in our samples, sources in clusters lie at $\rm z<1.2$. To make sure that our results and conclusions are not affected by the different redshift ranges probed by sources in different cosmic environments, we also restrict both our AGN and non-AGN datasets to $\rm z<1.2$. There are 551 (430 of which with spec-{\it{z}}) X-ray AGN and 16917 (6653 of which with spec-{\it{z}}) non-AGN galaxies that meet these two criteria (M$_*$, redshift). Among them, 31 (1048) AGN (non-AGN) are associated with clusters, 328 (9622) AGN (non-AGN) with filaments and 192 (6247) AGN (non-AGN) are in the field. The distribution of AGN in the L$_X-$redshift plane is presented in Fig. \ref{fig_lx_redz}. These sources are used in the analysis presented in Sect. 4.1. 

\begin{figure}
\centering
  \includegraphics[width=0.49\linewidth, height=4.5cm]{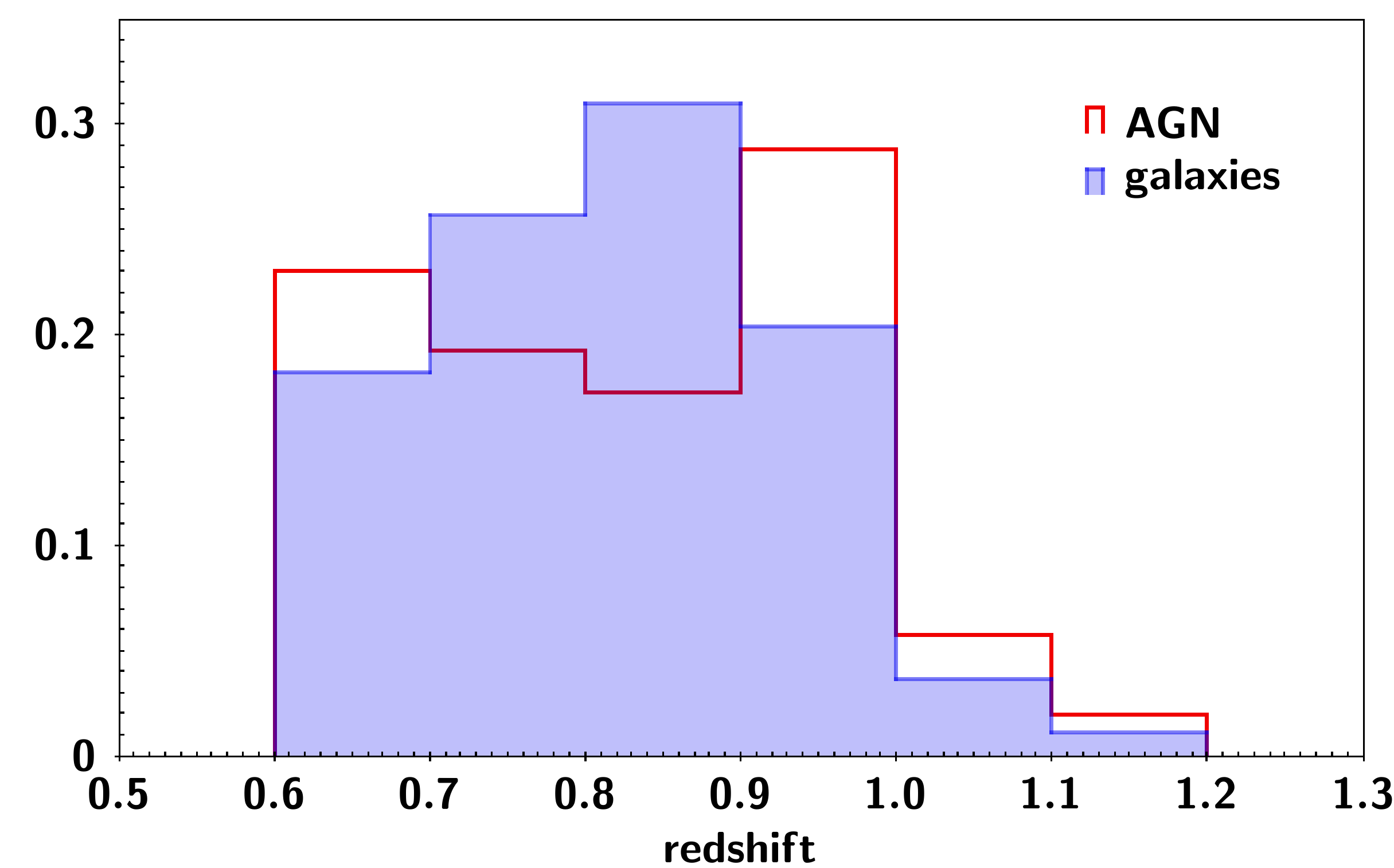}
  \includegraphics[width=0.49\linewidth, height=4.5cm]{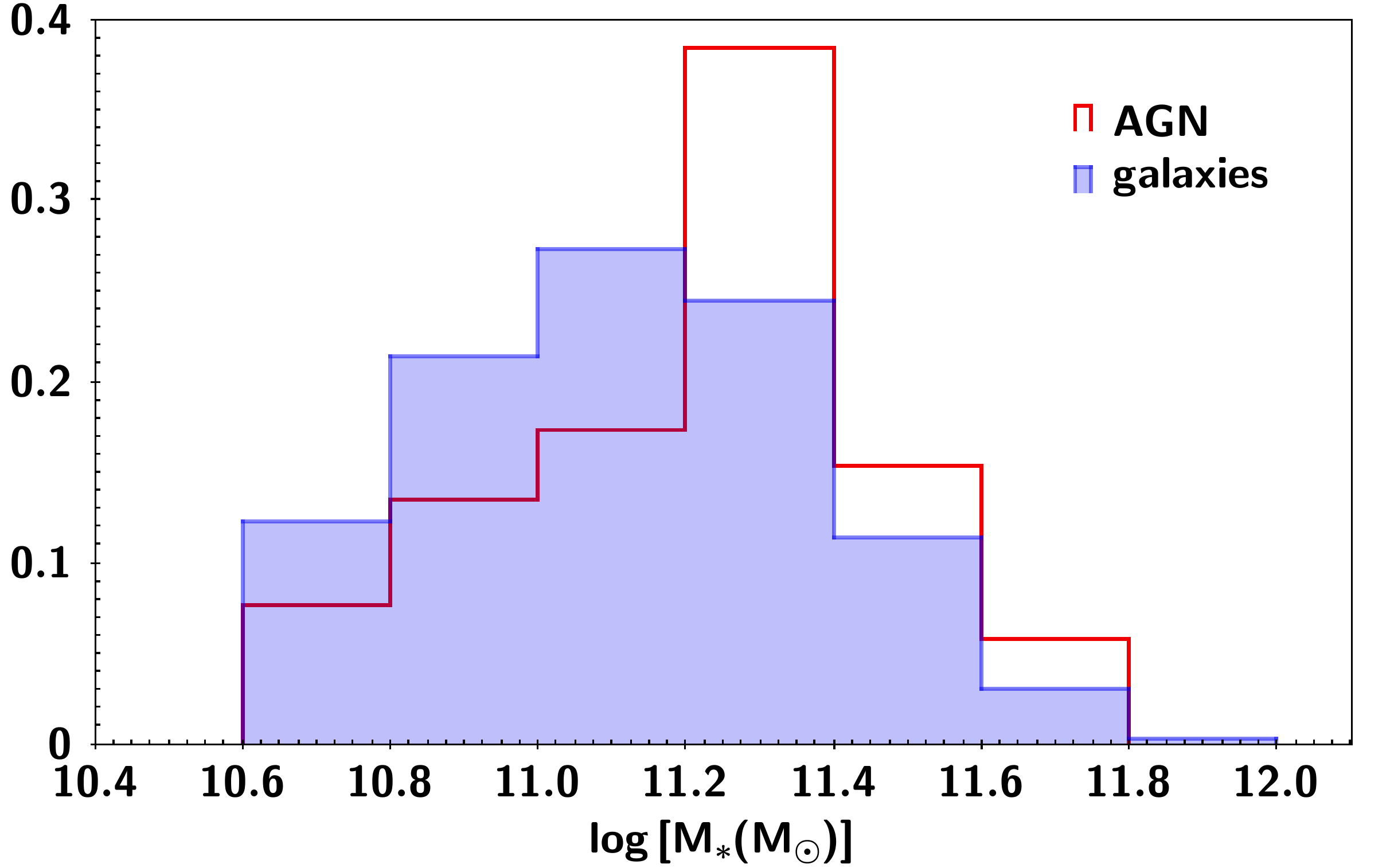}
  \caption{Distributions of galaxy properties. The left panel shows the distributions of redshift and the right panel of M$_*$, for the 52 AGN (red histograms) and 1172 non-AGN galaxies (blue shaded histograms), used to study the SFH of the two populations.}
  \label{fig_redz_mstar_distrib_sfh}
\end{figure} 

\subsection{Star-formation history}
\label{sect_sfh}

In \cite{Mountrichas2022c}, we examined the SFH of AGN and non-AGN galaxies and we found that AGN of low-to-moderate L$_X$ ($\rm log\,[L_{X,2-10keV}(erg\,s^{-1})]<44$) tend to have older stellar populations and are less likely to have experienced a recent star formation burst compared to non-AGN systems. Here, we study the SFH of the two populations, but, in addition, we take into account their cosmic environment (Sect. 4.2). For that purpose, we cross-match our AGN and galaxies in the control sample with the LEGA-C catalogue \citep{Wel2016, Straatman2018, Wel2021}. This catalogue includes spectra for 4081 galaxies (3741 unique objecs) within $\rm 0.6<z<1.3$. We make use of two stellar-age-sensitive tracers that are available in the LEGA-C dataset. These are the equivalent width (EW) of H$_\delta$ absorption, and the D$_n$4000 index \citep[e.g.,][]{Kauffmann2003a,Wu2018}. D$_n$4000 is small for young stellar populations and large for old, metal-rich galaxies. On the other hand, the EW of H$_\delta$ rises rapidly in the first few hundred million years after a burst of star formation, when O- and B-type stars dominate the spectrum, and then decreases when A-type stars fade \citep[e.g.,][]{Kauffmann2003a, Wu2018}.

We apply the same quality selection criteria described in Sect. 2.2.3 in \cite{Mountrichas2022c} that reduce the number of sources in the LEGA-C catalogue to 2834. Then, we cross match our X-ray and galaxy control samples with the LEGA-C dataset. This results in 52 AGN and 1,172 non-AGN galaxies. The average uncertainties on H$_\delta$ and D$_n$4000 are 17\% and 3\%, respectively. We note that the control sample has been reduced to those sources with $\rm log\,[M_*(M_\odot)]>10.6$ to better match the M$_*$ range of the X-ray dataset. The redshift and M$_*$ distributions of the two populations are shown in Fig. \ref{fig_redz_mstar_distrib_sfh}. Both distributions are similar between the two populations. A Kolmogorov-Smirnov (KS) test gives a p-value of 0.57 and 0.12 for the redshift and M$_*$ distributions of AGN and non-AGN galaxies, respectively (two distributions differ with a statistical significance of $\sim 2\sigma$ for a p-value of 0.05). Similar values are found applying a Mann-Whitney test (p-value of 0.90 and 0.29, respectively, for the redshift and M$_*$ distributions). Therefore, we consider that the M$_*$ and redshift distributions of AGN and non-AGN galaxies are not significantly different to affect the results presented in Sect. 4.2. Nevertheless, we confirm that if we account for the (small) differences in the redshift and M$_*$ distributions of the two populations, following the methodology described in e.g., \cite{Mountrichas2019, Masoura2021, Buat2021, Mountrichas2022c} (i.e., weigh the sources based on their M$_*$ and redshift), it does not change our results and overall conclusions. 

\subsection{Morphology}

To examine the role of morphology of AGN and non-AGN galaxies in different cosmic environments (see Sect 4.3), we use the catalogue presented in \cite{Ni2021}. The morphological classification was done using a deep-learning-based method to separate sources into BD and non-BD galaxies. The term BD refers to galaxies that only display a significant spheroidal component, without obvious disc-like or irregular components \citep[for more information, see Appendix C and Sect. 2.3 in][respectively]{Ni2021, Yang2019}.

We cross-match the datasets described in Sect. \ref{sect_sfh} with the catalogue of \cite{Ni2021}, which results in 37 AGN and 1086 non-AGN galaxies. Type 1 AGN have been removed from the \cite{Ni2021} sample, since they can potentially affect host galaxy morphological measurements (see their Sect. 2.4). We also note, that the redshift and M$_*$ distributions of AGN and non-AGN galaxies used in this part of our analysis, are very similar to those presented in Fig. \ref{fig_redz_mstar_distrib_sfh} and thus we assume that the small differences in the redshift and M$_*$ distributions of the two populations, do not affect our results.

\section{Results}
\label{sec_results}

In this section, we compare the SFR of AGN and non-AGN galaxies as a function of the L$_X$, for different cosmic environments. We also examine if the results of this comparison are affected by the exclusion of quiescent systems from both populations. Furthermore, we study the SFH of the two population for different density fields. Finally, we compare the effect of morphology and environment on the stellar populations of AGN and non-AGN systems.

\begin{figure}
\centering
  \includegraphics[width=0.9\linewidth, height=7cm]{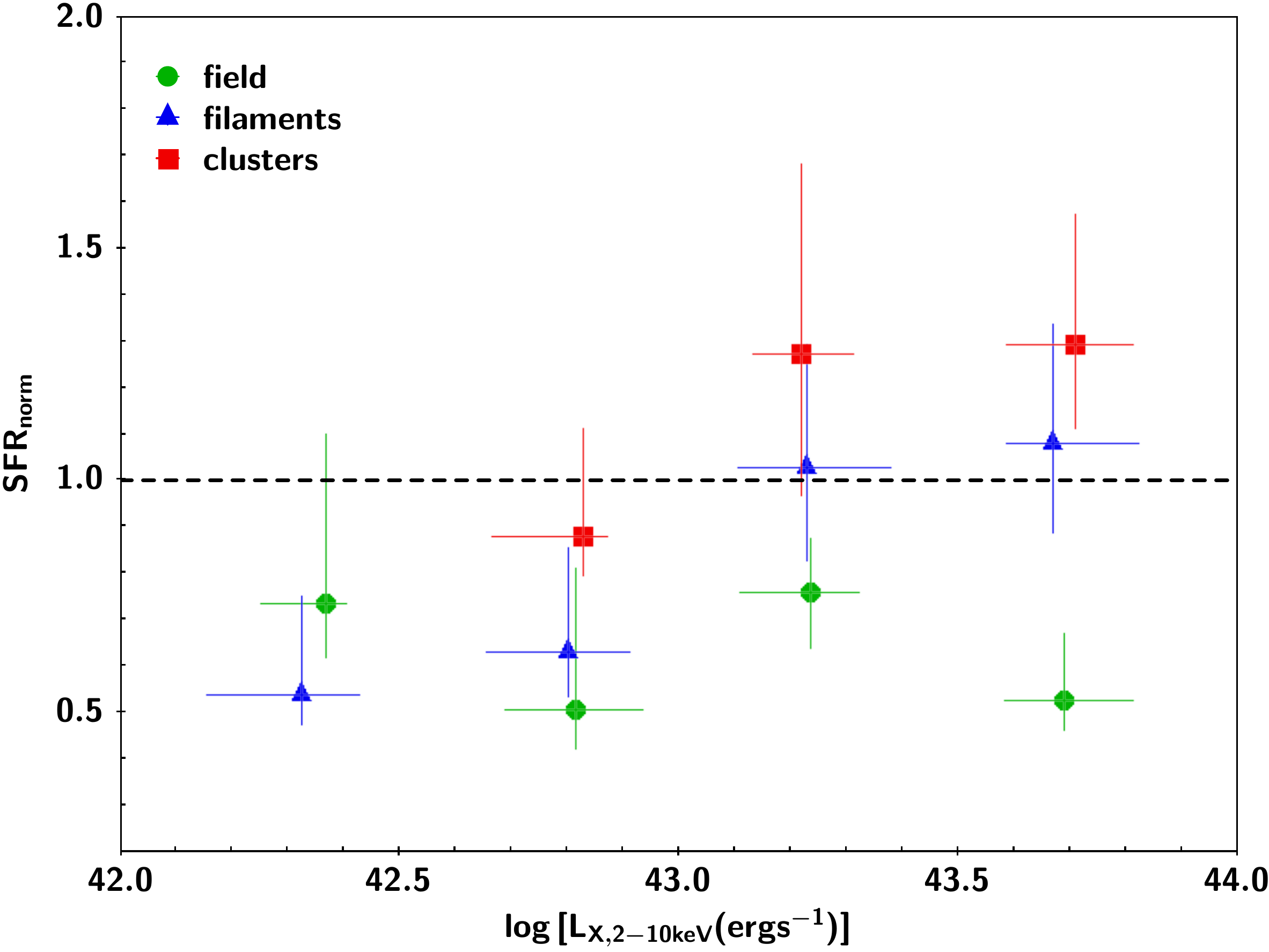}
  \includegraphics[width=0.9\linewidth, height=7cm]{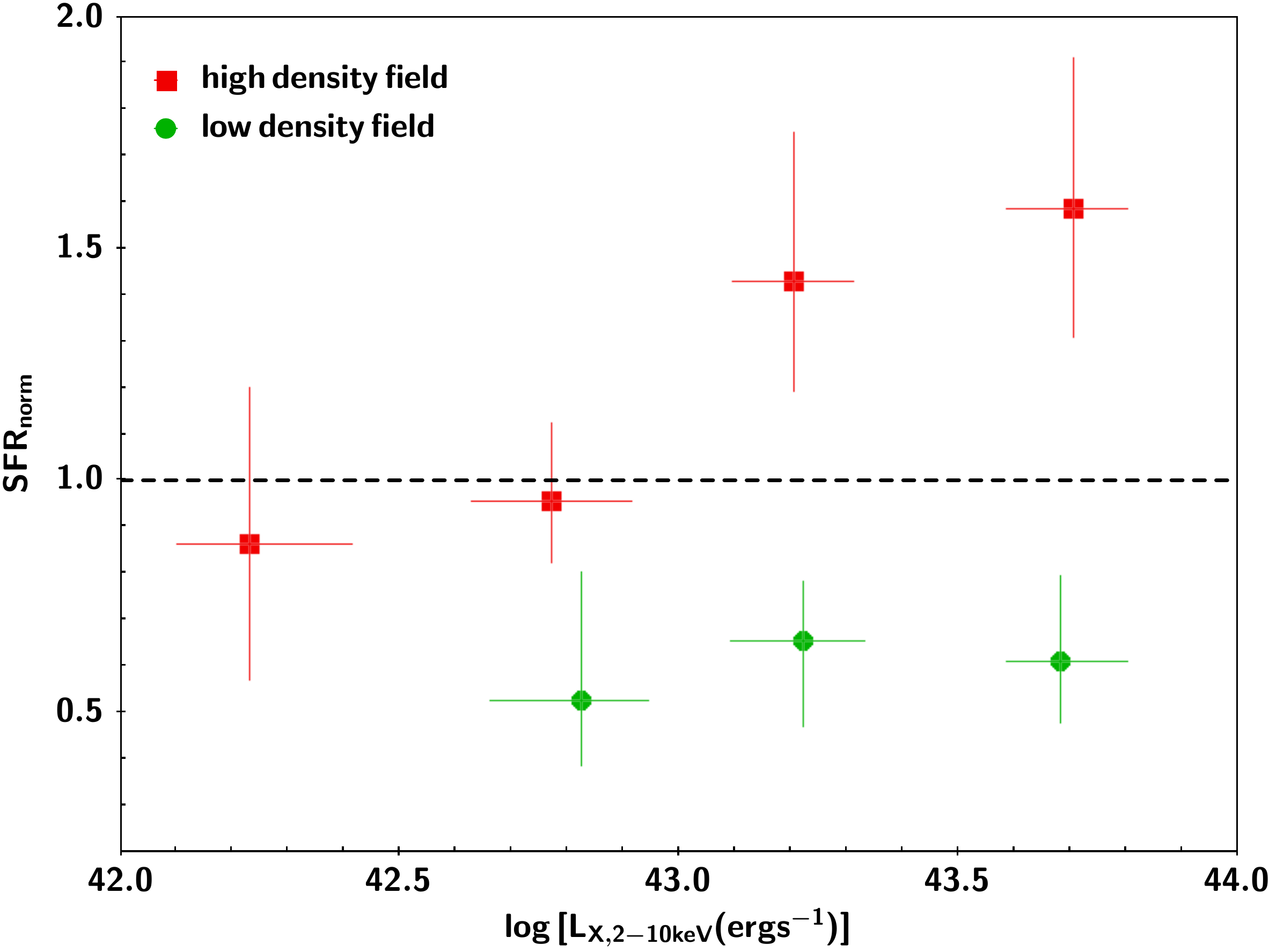}
    \includegraphics[width=0.9\linewidth, height=7cm]{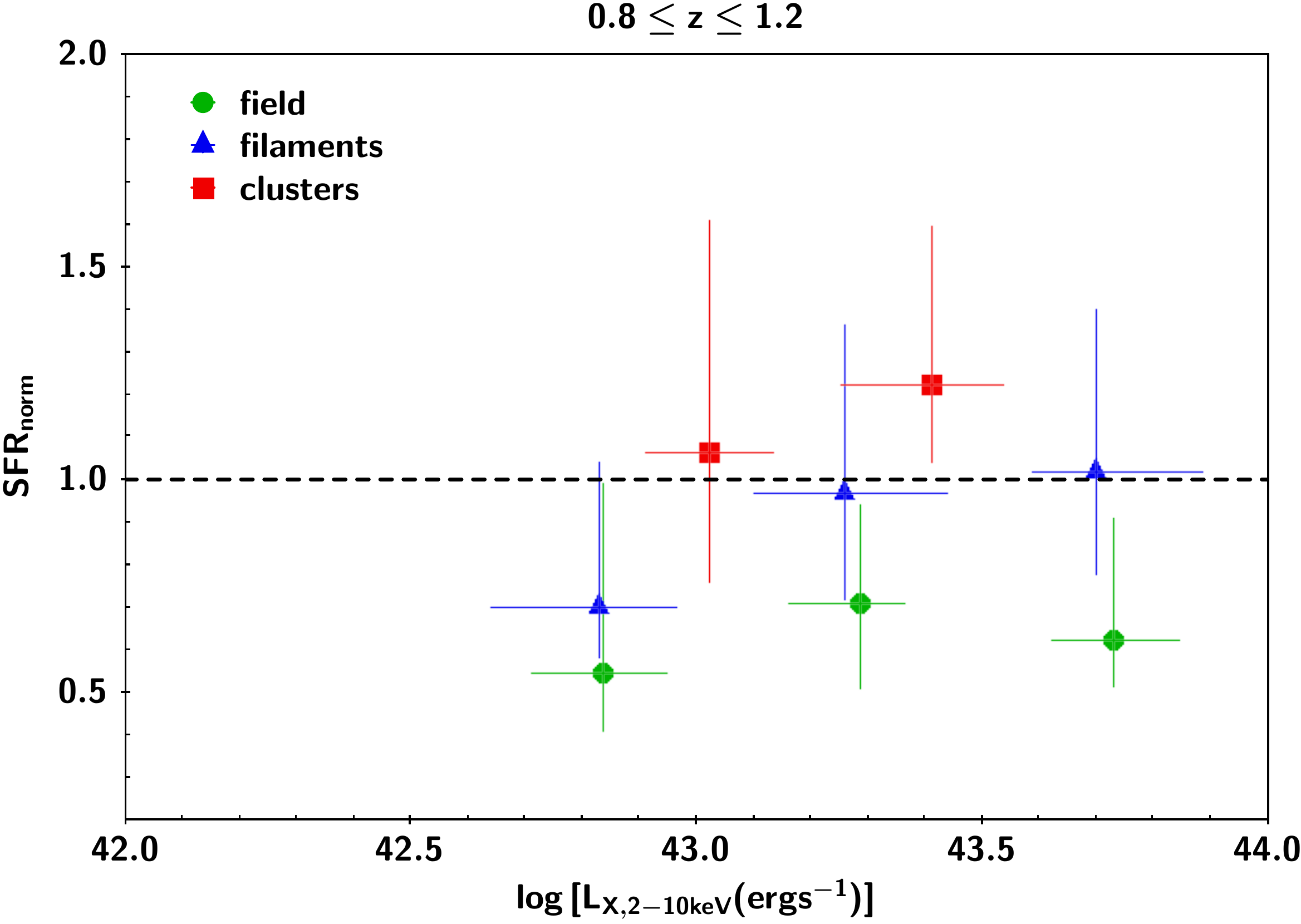}
  \caption{SFR$_{norm}$ ($\rm =\frac{SFR_{AGN}}{SFR_{non-AGN}}$) as a function of L$_X$. Median values are presented. The measurements are grouped in bins of L$_X$, of 0.5\,dex width. The errors presented are $1\,\sigma$, calculated using bootstrap resampling. The top panel presents the results when we classify sources based on their cosmic environment, as indicated in the legend. The middle panel shows the results when sources are classified using their overdensity values (see text for more details). The bottom panel is similar to the top panel, but we have restricted our dataset to $\rm 0.8\leq z\leq 1.2$.}
  \label{fig_sfrnorm_lx}
\end{figure} 

\begin{figure}
\centering
  \includegraphics[width=0.9\linewidth, height=7cm]{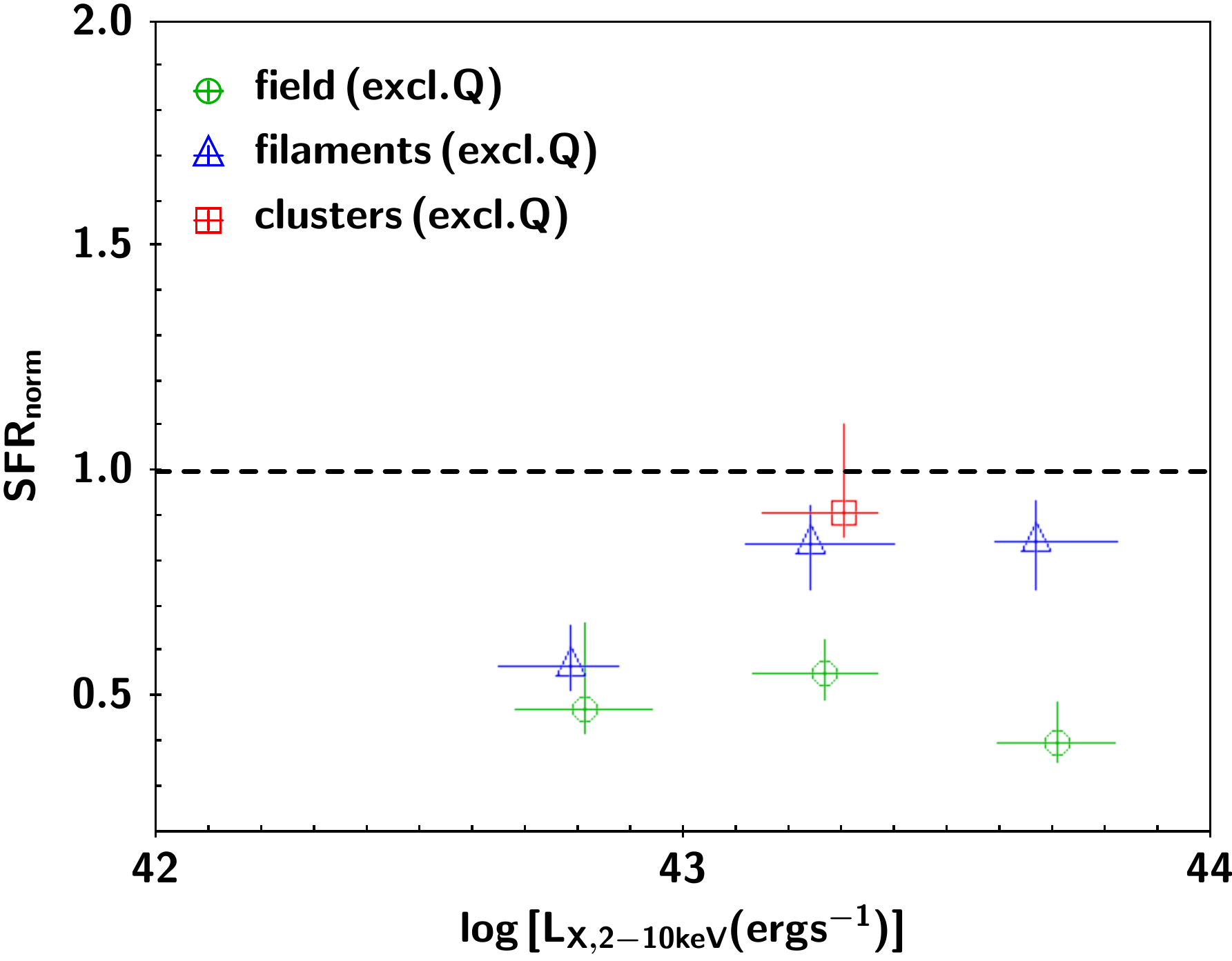}
  \includegraphics[width=0.9\linewidth, height=7cm]{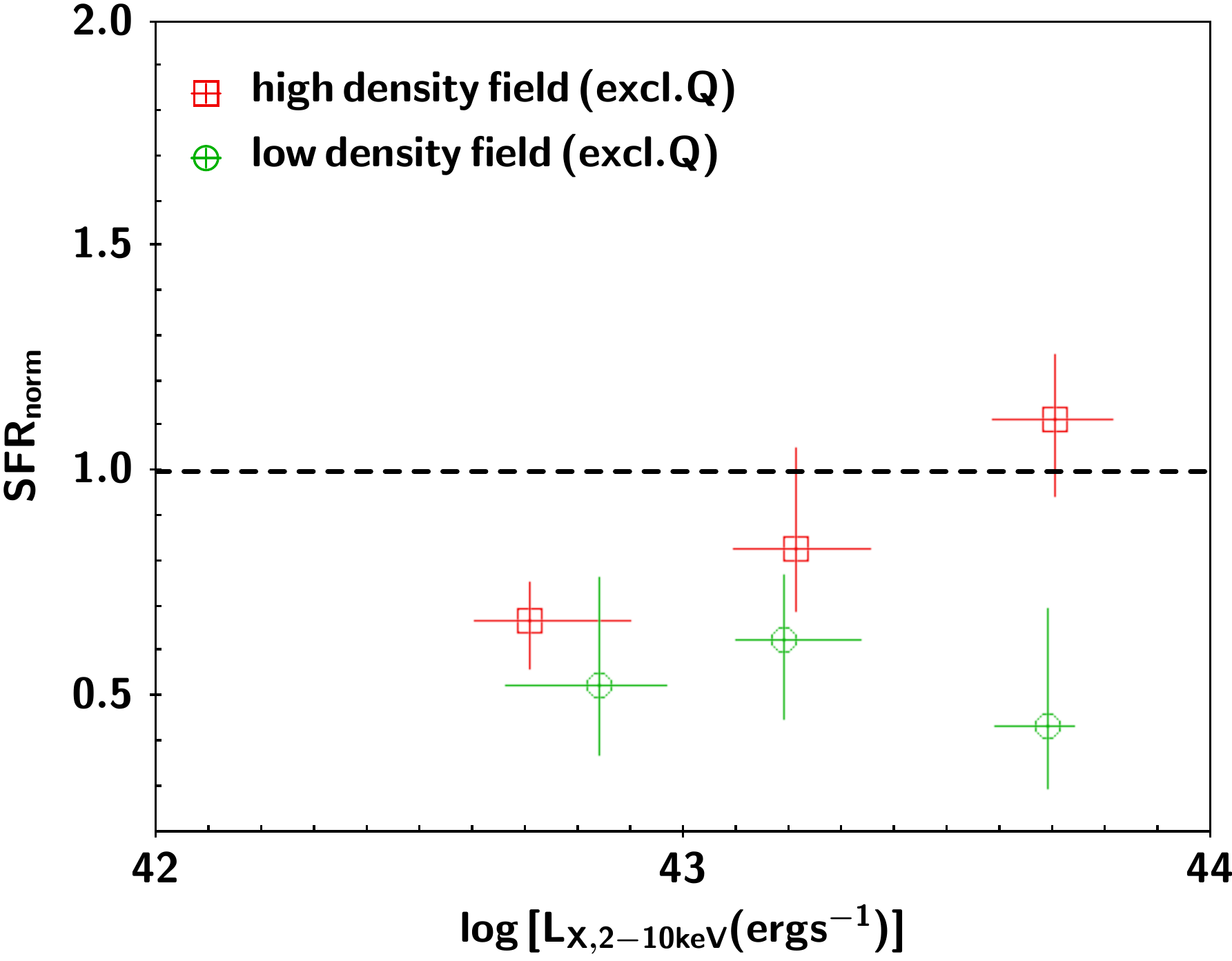}
  \caption{Same as Fig. \ref{fig_sfrnorm_lx}, but excluding quiescent (Q) systems both from the AGN and galaxy samples.}
  \label{fig_sfrnorm_lx_excl_q}
\end{figure}

\subsection{Comparison of the SFR of AGN and non-AGN galaxies, as a function of L$_X$ and cosmic environment}

First, we compare the SFR of AGN host galaxies with the SFR of non-AGN systems, as a function of the cosmic environment. For that purpose, we use the SFR$_{norm}$ parameter. SFR$_{norm}$ is defined as the ratio of the SFR of galaxies that host AGN to the SFR of non-AGN systems, with similar M$_*$ and redshift \citep[e.g.,][]{Mullaney2015, Masoura2018, Bernhard2019, Masoura2021, Mountrichas2021c, Mountrichas2022a, Mountrichas2022b}. For the estimation of SFR$_{norm}$ we use the X-ray and galaxy control samples described in Sect. \ref{sec_data}. SFR$_{norm}$ is measured following the process of our previous studies \citep[e.g.,][]{Mountrichas2021c, Mountrichas2022a, Mountrichas2022b}. Specifically, the SFR of each X-ray AGN is divided by the SFR of galaxies in the control sample that are within $\pm 0.2$\,dex in M$_*$, $\rm \pm 0.075\times (1+z)$ in redshift and are in similar density fields. Furthermore, each source is weighted based on the uncertainty of the SFR and M$_*$ measurements made by CIGALE. Then, the median values of these ratios are used as the SFR$_{norm}$ of each X-ray AGN. We note that our measurements are not sensitive to the choice of the box size around the AGN. Selecting smaller boxes, though, has an effect on the errors of the calculations \citep{Mountrichas2021c}.

The top panel of Fig. \ref{fig_sfrnorm_lx} presents the SFR$_{norm}$ as a function of L$_X$, for different cosmic environments. Measurements are the median values of SFR$_{norm}$ grouped in bins of L$_X$ of size 0.5\,dex. The errors presented are $1\,\sigma$, calculated using bootstrap resampling \citep[e.g.,][]{Loh2008}. Only bins that have more than 10 sources are presented. We notice that the SFR$_{norm}$-L$_X$ relation is flat for sources found in the field. However, an increase of SFR$_{norm}$ is observed, in particular at $\rm log\,[L_{X,2-10keV}(erg\,s^{-1})]>43$, for sources associated with filaments and clusters. This implies that for X-ray luminosities probed by our dataset, AGN in the field have lower SFR compared to the SFR of non-AGN systems, regardless of the L$_X$. However, in denser environments, the SFR of AGN compared to non-AGN galaxies depends on the L$_X$. At $\rm log\,[L_{X,2-10keV}(erg\,s^{-1})]<43$, AGN appear to have lower SFR compared to non-AGN, wheras at moderate L$_X$, AGN have similar or larger SFR compared to non-AGN systems. Previous studies \citep{Mountrichas2022a, Mountrichas2022b}, found that the SFR of low to moderate L$_X$ AGN (i.e., $\rm log\,[L_{X,2-10keV}(erg\,s^{-1})]<44$) is lower, or at most equal to the SFR of non-AGN systems. In light of our current measurements, these previous results could be attributed to the density field of the sources. That is, low to moderate L$_X$ AGN have lower SFR compared to non-AGN systems when both populations are found in the field, but the SFR of (moderate L$_X$) X-ray AGN is similar or larger to that of non-AGN galaxies when the sources are associated with filaments or clusters.

We repeat the same exercise, but we now classify AGN and non-AGN using their overdensity values. Specifically, those sources that have overdensity values that belong to the highest 20\% of the total samples are classified as sources in `high density fields', whereas those with overdensity values that belong to the lowest 20\% of the total samples are characterized as sources in `low density fields'. Sources that belong to the `high density field' group have $\rm log(1+\delta)>0.28$ (both in the case of AGN and non-AGN galaxies), whereas those in the `low density field' group have $\rm log(1+\delta)<-0.12$. The SFR$_{norm}$ as a function of the L$_X$ for the two classes is presented in the middle panel of Fig. \ref{fig_sfrnorm_lx}. These results confirm our previous observations, but the trends are now more clear and appear statistical significant. Specifically, SFR$_{norm}$ in denser fields is higher compared to fields of low density, at a $\sim 2\sigma$ significance level and there is an increase of SFR$_{norm}$ with L$_X$ for sources in dense environments, whereas in fields of low density the SFR$_{norm}$-L$_X$ relation is flat and SFR$_{norm}$ is consistently lower than 1 (i.e., the SFR of AGN is lower compared to the SFR of non-AGN galaxies). 

%We note, that the same trends are observed when we split our AGN and non-AGN galaxies into two redshift bins, at $\rm z=0.9$. 

To examine if our results are affected by possible evolution with cosmic time within the redshift range that our datasets span ($\rm 0.3<z<1.2$), we restrict our samples to a narrow redshift interval, that is $\rm 0.8\leq z\leq 1.2$. There are 238 AGN and 8019 sources in the control sample within this redshift range. The results using these subsets are presented in the bottom panel of Fig. \ref{fig_sfrnorm_lx}. The statistical uncertainties are larger due to the smaller size of the subsets used, but the same trends are observed with those using the sources within the wider redshift interval.

Next, we examine if the trends observed above for the total AGN and non-AGN galaxies, hold in the case of star-forming sources, that is after excluding quiescent (Q) systems. In our previous studies \citep{Mountrichas2021c, Mountrichas2022a, Mountrichas2022b}, we have identified quiescent systems (both AGN and non-AGN galaxies), using their specific SFR measurements ($\rm sSFR=\frac{SFR}{M_*}$) and we have excluded them from our analysis, when studying the SFR$_{norm}$-L$_X$ relation. Therefore, we now use systems identified as quiescent in \cite{Mountrichas2022a} (see their Sect. 3.5) that studied the SFR$_{norm}$-L$_X$ in the COSMOS field, and we examine whether their fraction differs between AGN and non-AGN galaxies, in different cosmic environments. We find that the fraction of quiescent sources increases in denser fields for non-AGN galaxies, from $\sim 25\%$ for isolated galaxies (or galaxies in low density fields) to $\sim 40\%$ in the most dense environments. In the case of AGN, the fraction of quiescent systems is similar in all environments ($\sim 30\%$).

We, then, repeat the SFR$_{norm}$-L$_X$ measurements, excluding quiescent systems. The results are presented in Fig. \ref{fig_sfrnorm_lx_excl_q}. We note that due to the smaller size of these samples, there is only one bin that satisfies our requirement for the minimum number of sources ($>10$), in the case of AGN in clusters. Overall, when quiescent systems are not included in the analysis, the amplitude of SFR$_{norm}$ is lower (by $\sim 0.2$\,dex) for all cosmic environments and L$_X$ spanned by our datasets compared to the SFR$_{norm}$ values derived using the total samples (Fig. \ref{fig_sfrnorm_lx}). However, the trends mentioned earlier remain unchanged. Specifically, SFR$_{norm}$ is higher for denser environments and an increase of SFR$_{norm}$ with L$_X$ is observed for sources associated with filaments/clusters. 

\begin{figure}
\centering
  \includegraphics[width=0.85\columnwidth, height=6.2cm]{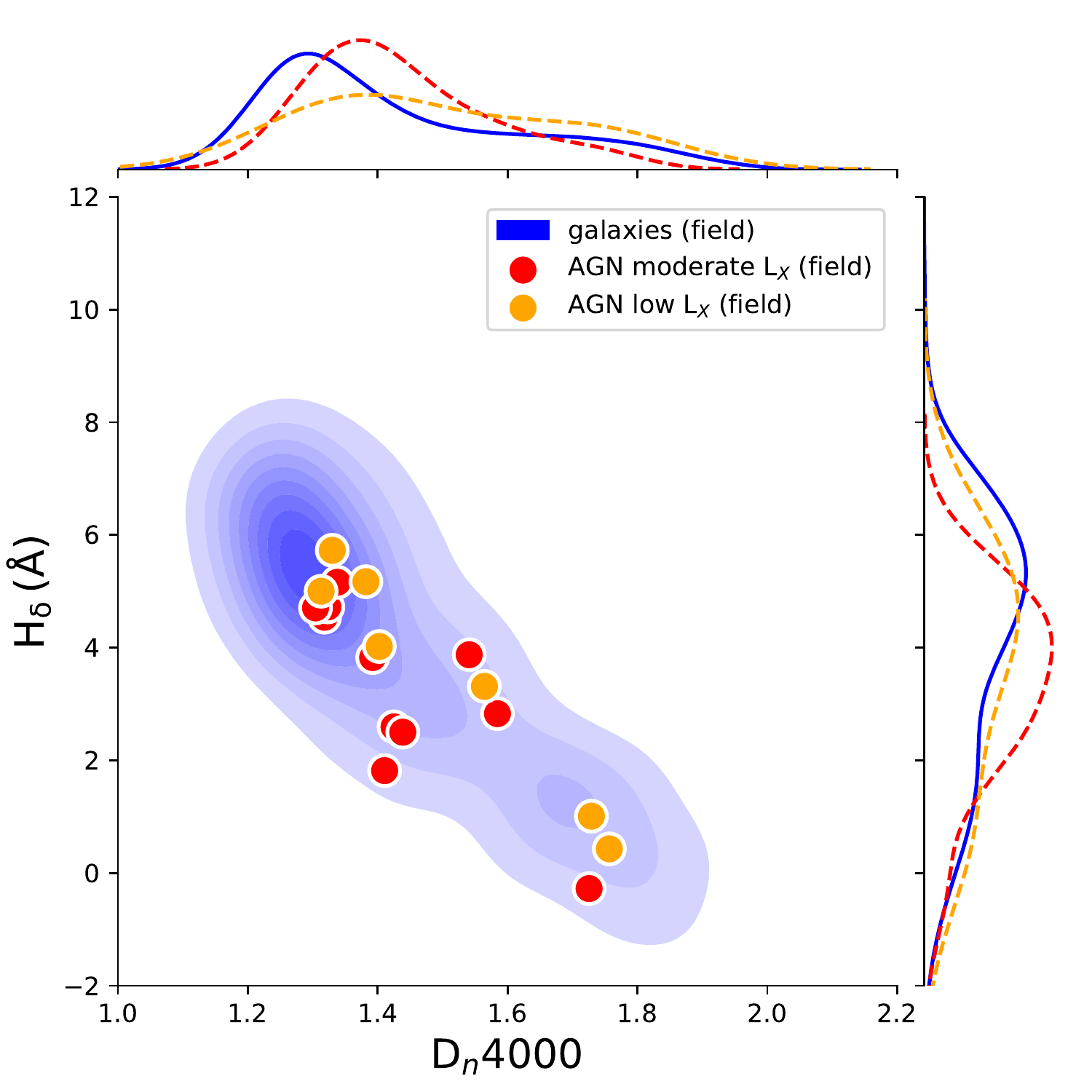}   
  \includegraphics[width=0.85\columnwidth, height=6.2cm]{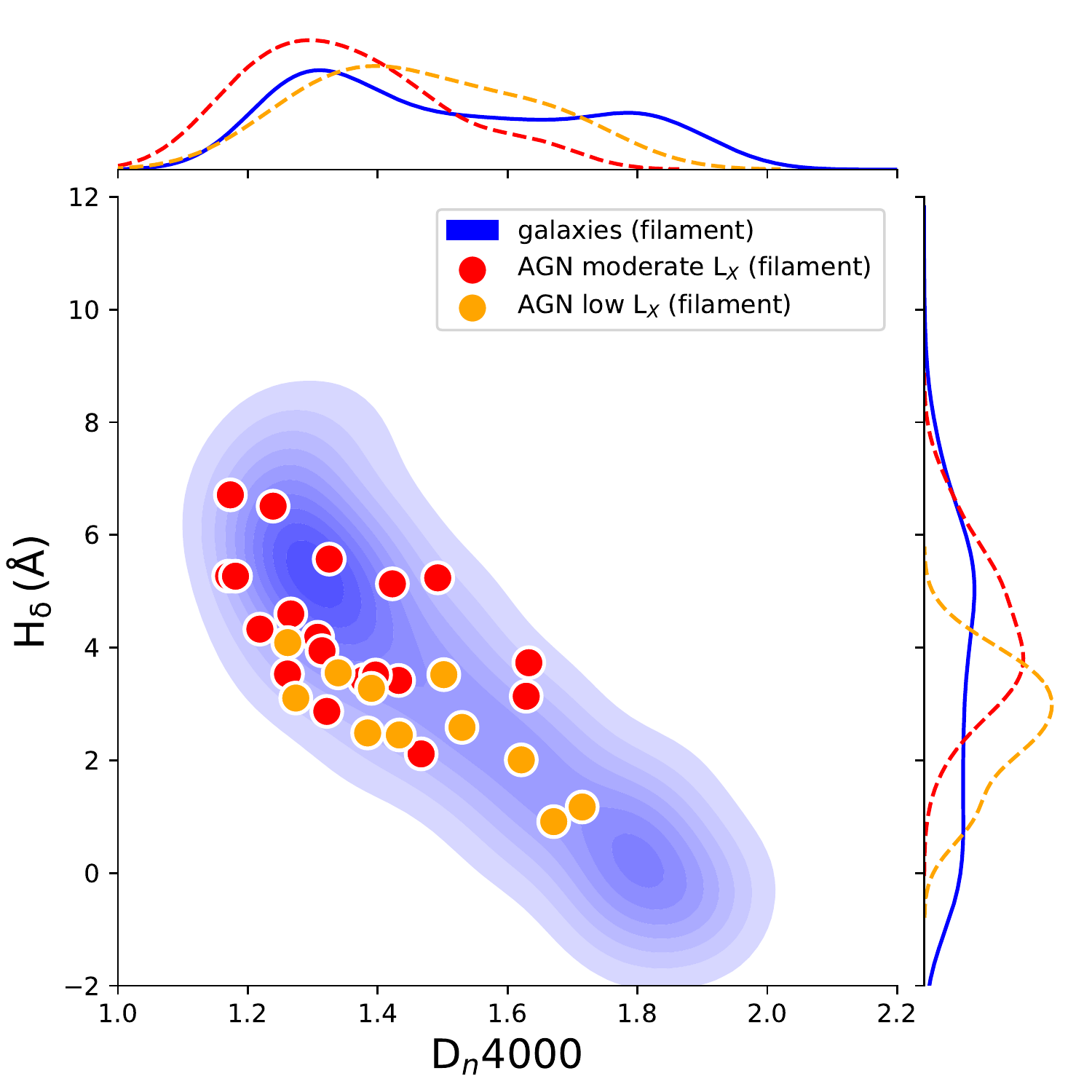} 
  \includegraphics[width=0.85\columnwidth, height=6.2cm]{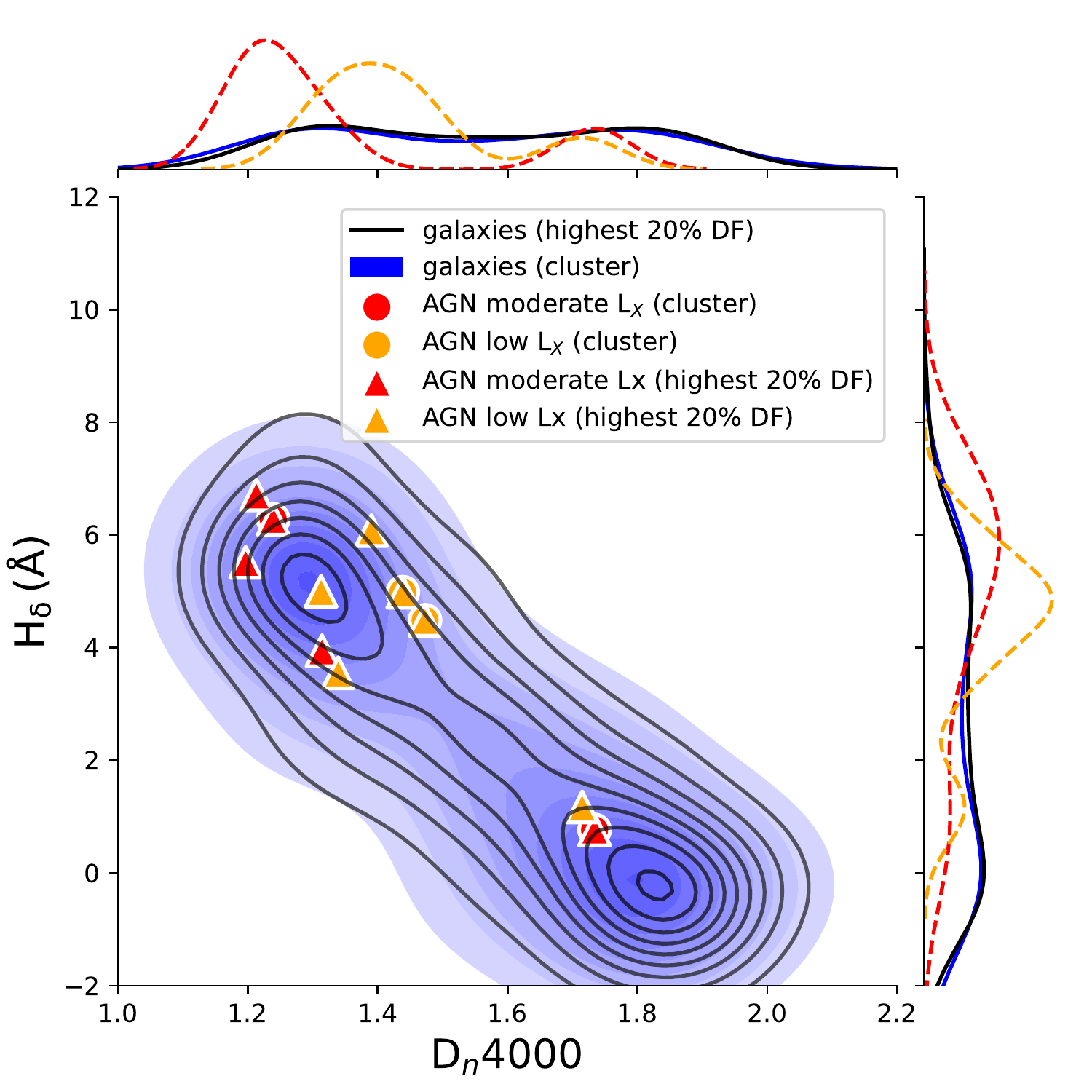}
  \caption{The distributions in the H$_\delta$-D$_n$4000 space of non-AGN galaxies (blue shaded contours) and low ($\rm log\,[L_{X,2-10keV}(erg\,s^{-1})]<43$; orange circles) and high ($\rm 43<log\,[L_{X,2-10keV}(erg\,s^{-1})]<44$; red circles) L$_X$ AGN, for different cosmic environments. The top panel, presents the distribution for isolated sources (field), the middle panel for sources associated with filaments and the bottom panel for sources found in the most dense fields (see text for more details).}
  \label{fig_sfh}
\end{figure}

\subsection{The star-formation histories of AGN and non-AGN galaxies, as a function of the cosmic environment}

In this section, we study the SFH of AGN and non-AGN galaxies. Our goal is to check whether their stellar populations agree with the picture drawn in Sect. 4.1 (e.g., systems with lower SFR are expected to have older stars compared to systems with higher SFR). Moreover, the study of the SFH could allow us to reveal what drives the trends we observed in the previous section. We note that the selection of sources with available D$_n$4000 and H$_\delta$ measurements does not bias our samples against Q systems. In other words, the AGN and non-AGN datasets used in this part of our analysis, include the same fraction of Q systems with the datasets used in the previous section. Therefore, a direct comparison can be made of the results of these two parts of our analysis.

Fig. \ref{fig_sfh} presents the distributions of AGN and non-AGN galaxies in the H$_\delta$-D$_n$4000 space, for sources in the field (top panel) and those associated with filaments (middle panel) and clusters (bottom panel). Prompted by the results in the previous section, we split our AGN into two L$_X$ bins, at $\rm log\,[L_{X,2-10keV}(erg\,s^{-1})]=43$. Red circles indicate AGN with $\rm log\,[L_{X,2-10keV}(erg\,s^{-1})]>43$ (`moderate L$_X$') and orange circles represent AGN with $\rm log\,[L_{X,2-10keV}(erg\,s^{-1})]<43$ (`low L$_X$'). The number of AGN and non-AGN available are presented in Table \ref{table_sfh_numbers}.

Regarding, the D$_n$4000 and H$_\delta$ distributions for AGN and non-AGN galaxies that are in the field (top panel of Fig. \ref{fig_sfh}), we notice that the two AGN populations have similar stellar ages, which are higher compared to non-AGN systems. A KS test yields a p-value of $\sim 1$ for the D$_n$4000 distributions of low and moderate L$_X$ AGN. Comparing the distributions between AGN and non-AGN galaxies, we get a p-value of 0.20 and 0.008 for the D$_n$4000 distributions of moderate and low L$_X$ AGN compared to non-AGN systems, that indicates a statistically significant difference, at least in the case of low L$_X$ AGN compared to non-AGN. AGN, also, tend to have lower H$_\delta$ values compared to sources in the control sample. However, based on a KS test, this difference does not appear statistically significant. Merging the two AGN populations and comparing their H$_\delta$ and D$_n$4000 distributions with those of their non-AGN counterparts does not affect the overall results. Specifically, AGN appear to have higher D$_n$4000 and lower H$_\delta$ values compared to non-AGN galaxies. Based on the p-values of the KS tests, the first difference is statistically significant (p-value of 0.02 and 0.05, respectively, for D$_n$4000 and H$_\delta$). 

The above suggest that AGN that live in the field tend to have older stellar populations and are less likely to have experienced a recent star-formation burst compared to isolated non-AGN systems. These results are in agreement with those presented in Fig. \ref{fig_sfrnorm_lx}. The picture that emerges is that AGN that are in the field have similar star-formation rates and histories regardless of the AGN power (L$_X$) and have, on average, lower SFR and older stellar ages compared to their non-AGN field counterparts.  

On the other hand, as shown in the middle panel of Fig. \ref{fig_sfh}, AGN that live in denser environments (i.e., filaments) have different star-formation histories, depending on their L$_X$. Specifically, AGN with $\rm log\,[L_{X,2-10keV}(erg\,s^{-1})]>43$ tend to have lower D$_n$4000 ($=1.32$ vs. $= 1.44$) and higher H$_\delta$ values ($= 4.20$ vs. $= 2.56$), compared to their lower L$_X$ counterparts (p-value of $\sim 0.3$, for both spectral indices). These differences, although not statistically significant based on KS tests, are in agreement with the results we found in the previous section, where higher L$_X$ AGN have increased SFR$_{norm}$ compared to lower L$_X$ sources. Regarding the non-AGN galaxies, these appear to have flatter D$_n$4000 and H$_\delta$ distributions compared to AGN (median D$_n4000 =1.46$, H$_\delta =3.46$).  

\begin{table*}
\caption{Number of X-ray AGN and sources in the control sample with available D$_n$4000 and H$_\delta$ measurements from the LEGA-C catalogue, in different cosmic environments. Median D$_n$4000 and H$_\delta$ values of each subset is also presented.}
\centering
\setlength{\tabcolsep}{3mm}
\begin{tabular}{ccccccc}
 \hline
 & \multicolumn{2}{c}{field} &  \multicolumn{2}{c}{filament} &  \multicolumn{2}{c}{cluster} \\
 \hline
L$_X$ (erg/s) & $<10^{43}$ & $\geq 10^{43}$ & $<10^{43}$ & $\geq 10^{43}$ & $<10^{43}$ & $\geq 10^{43}$ \\
 \hline
no. AGN & 7 & 11 & 11 & 19 & 2 (6) & 2 (5) \\
D$_n4000$, H$_\delta$ (AGN) & 1.40, 4.00 & 1.41, 3.89 & 1.44, 2.56 & 1.32, 4.20 & 1.44, 4.95 & 1.22, 5.53 \\ 
no. galaxies & \multicolumn{2}{c}{490} & \multicolumn{2}{c}{603} & \multicolumn{2}{c}{79 (244)}  \\
D$_n4000$, H$_\delta$ (galaxies) & \multicolumn{2}{c}{1.36, 4.45} & \multicolumn{2}{c}{1.46, 3.46} & \multicolumn{2}{c}{1.55, 2.23}  \\
  \hline
\label{table_sfh_numbers}
\end{tabular}
\tablefoot{Numbers in the parentheses show the number of AGN when we classify sources using their overdensity values (see text for more details).}
\end{table*}

\begin{table*}
\caption{Median D$_n$4000 and H$_\delta$ values for star-forming (SF) and quiescent (Q) AGN and non-AGN galaxies in different density fields.}
\centering
\setlength{\tabcolsep}{3mm}
\begin{tabular}{ccccc}
 \hline
 & \multicolumn{2}{c}{low density field} &  \multicolumn{2}{c}{high density field} \\
 \hline
  & SF & Q & SF & Q  \\
 \hline
D$_n4000$, H$_\delta$ (AGN) & 1.31, 4.18 & 1.44, 0.27 & 1.34, 4.98 & 1.72, 1.17  \\ 
D$_n4000$, H$_\delta$ (galaxies) & 1.30, 5.13 & 1.67, 1.74 & 1.35, 1.78 & 1.78, 0.21  \\
  \hline
\label{table_sfh_numbers_Q}
\end{tabular}
\end{table*}

\begin{table*}
\caption{Number of X-ray AGN and sources in the control sample for different morphologies and cosmic environments. The median D$_n$4000 and H$_\delta$ values of each subset is also presented.}
\centering
\setlength{\tabcolsep}{3mm}
\begin{tabular}{ccccccc}
 \hline
 & \multicolumn{2}{c}{field} &  \multicolumn{2}{c}{filament} &  \multicolumn{2}{c}{cluster} \\
 \hline
morphology & BD & non-BD & BD & non-BD & BD & non-BD \\
no. AGN & \multicolumn{2}{c}{12 (32\%)} & \multicolumn{2}{c} {21 (57\%)} & \multicolumn{2}{c}{4 (11\%) [13]} \\
AGN & 6(50\%) & 6(50\%) & 3(14\%) & 18(86\%) & 0(0\%) [0(0\%)] & 4(100\%) [13(100\%)] \\
D$_n4000$, H$_\delta$ (AGN) & 1.58, 5.00 & 1.40, 4.54 & 1.63, 2.58 & 1.42, 3.43 & nan, nan & 1.47, 4.98 \\ 
no. galaxies & \multicolumn{2}{c}{494(41\%)} & \multicolumn{2}{c}{560(53\%)} & \multicolumn{2}{c}{72(7\%) [218]}  \\
galaxies & 87(19\%) & 367(81\%) & 121(22\%) & 439(78\%) & 20(28\%) [62(28\%)] & 52(72\%) [156(72\%)] \\
D$_n4000$, H$_\delta$ (galaxies) & 1.67, 1.23 & 1.32, 5.03 & 1.76, 0.64 & 1.38, 4.19 & 1.80, -0.04 & 1.40, 3.62 \\ 
  \hline
\label{table_morphology}
\end{tabular}
\tablefoot{Numbers in the square brackets show the number of AGN when we classify sources using their overdensity values (see text for more details).}
\end{table*}

Finally, the bottom panel of Fig. \ref{fig_sfh}, presents the  D$_n$4000 and H$_\delta$ distributions for AGN and non-AGN systems that are associated with the most dense cosmic environments. Since there are only four AGN that live in clusters in our dataset, we also follow the same approach we followed in the previous section. That is, we also classify sources based on their overdensity values. We notice that the results are very similar, both for the D$_n$4000 and the H$_\delta$, for the non-AGN galaxies, regardless of whether the classification is based on the cosmic environment (clusters) or the the overdensity value (i.e., the solid black and red lines). AGN with $\rm log\,[L_{X,2-10keV}(erg\,s^{-1})]>43$ have lower D$_n$4000 and higher H$_\delta$ values compared to their lower L$_X$ counterparts (p-value of $0.00003$ and 0.004, for D$_n$4000 and H$_\delta$, respectively) and the non-AGN galaxies (p-value of $0.02$ and 0.006, for D$_n$4000 and H$_\delta$, respectively). Based on the p-values of the KS-tests, these differences are statistically significant at a level of $>2\sigma$. This confirms our findings, presented in the previous section, for higher SFR of high L$_X$ AGN compared to lower L$_X$ X-ray sources and non-AGN systems. 

Table \ref{table_sfh_numbers} shows the median D$_n$4000 and H$_\delta$ values for the different populations and cosmic environments. Non-AGN galaxies tend to have higher D$_n$4000 and lower H$_\delta$ values as we move to denser fields, which indicate that their stellar population becomes older in denser environments, in agreement with most previous studies \citep[e.g.,][]{PerezMillan2023}. In addition to the median values presented in Table \ref{table_sfh_numbers}, there are noteworthy differences in the D$_n$4000 and H$_\delta$ distributions of galaxies in different cosmic environments (Fig. \ref{fig_sfh}). For galaxies in the field, the two distributions, present a long tail. This tail becomes more prominent for sources associated with filaments and in the case of galaxies in the densest environments the distributions become flat. We confirm that these long tails/secondary peaks are populated, mainly, by Q galaxies and their prominence is associated with the increased fraction of Q non-AGN systems in denser environments.

On the other hand, AGN and in particular X-ray sources with $\rm log\,[L_{X,2-10keV}(erg\,s^{-1})]<43$ have consistent H$_\delta$ and, especially, D$_n$4000 values in all cosmic environments (Table \ref{table_sfh_numbers}) and their D$_n$4000 and H$_\delta$ distributions have less prominent tails compared to non-AGN galaxies (Fig. \ref{fig_sfh}). More luminous AGN show a tendency of lower D$_n$4000 and higher H$_\delta$ values in the most dense environments, although the sub-sample is small and larger datasets are required to confirm these findings. 

These results can provide an explanation as to what drives the trends we observed in the previous section. Specifically, the increase of the SFR$_{norm}$ as we move from the field to filaments and to clusters (Fig. \ref{fig_sfrnorm_lx}) is mainly driven by the lower SFR of galaxies in higher density fields and to a lesser degree due to the increased SFR of the more luminous AGN in denser cosmic environments. 

%However, this picture reverses at the most dense cosmic environments and this is, mostly, driven by the higher stellar ages of galaxies in the most dense fields. In other words, the increase of the SFR$_{norm}$ as we move from the field to filaments and to clusters, that we detected in the previous section, is mainly driven by the lower SFR of galaxies in the denser fields and to a lesser degree due to the increased SFR of the more luminous AGN in denser cosmic environments.

We repeat the same exercise, separating now the sources into SF and Q systems. The results are shown in Table \ref{table_sfh_numbers_Q}. Due to the small number of available AGN we do not split the AGN based on their L$_X$. We also characterize the cosmic environment of sources based on their overdensity values (see Sect. 4.1). The results show that the stellar populations of Q systems, both in the case of AGN and non-AGN galaxies, depend on the cosmic environment, whereas for SF systems the differences with the density field are smaller. We note, however, that in the case of AGN, the measurements come from a small number of sources in each density field. Specifically, both at high and low density fields, our AGN subsamples include eight SF and three Q systems.

\subsection{Environment vs. morphology}

In this section, we examine the morphology of AGN and non-AGN galaxies, in different cosmic environments. This will allow us to study the SFH of AGN and non-AGN systems in different environments, at fixed morphological type. It will also enable us to examine the link between morphology (e.g., BD) and galaxy phase (e.g., Q) for the two populations. 

There are nine (24\%) AGN in our dataset that live in BD systems. A similar fraction (21\%) of BD galaxies is also found in the non-AGN sample. We also note that the fraction of Q systems is similar ($\sim 40\%$) for AGN and non-AGN galaxies. We find  similar fractions between the AGN and the galaxies in the control sample to be associated with fields, filaments and clusters (see Table \ref{table_morphology}). A noteworthy difference, though, is that the percentage of BD systems shows an (mild) increase towards higher density environments among the non-AGN galaxies, similar to the increase of the fraction of Q systems in denser fields. On the contrary, the percentage of BD systems that also host AGN is significantly reduced in AGN associated with dense cosmic environments, while the fraction of Q AGN hosts remains roughly the same, independently of the density field. When we associate sources with high density regions based on the overdensity values of the sources, the fractions of BD systems in both the AGN and the non-AGN datasets (numbers in square brackets in Table \ref{table_morphology}) is the same as those we found using their cosmic environment, and therefore the same conclusions are reached. 

Therefore, based on our analysis, BD systems are nearly equally likely to be found in non-AGN galaxies regardless of their cosmic environment. In the case of galaxies that host (type 2) AGN, BD systems are preferentially found in AGN associated with less dense fields. Moreover, the link between morphology (BD) and galaxy phase (Q) is weaker for AGN compared to non-AGN galaxies. This is confirmed by the study of the stellar populations, presented in Table \ref{table_morphology}. The stellar population of non-AGN galaxies is different for BD and non-BD systems, as opposed to galaxies that host AGN, in which case smaller differences are observed in the stellar populations for different morphologies \citep[see also][]{Mountrichas2022c}.  We note, however, that in the case of AGN the size of the available dataset is small and larger samples are required to confirm our findings.

%Nevertheless, even in the case of AGN, and certainly in the case of non-AGN galaxies, morphology is a more differentiating factor of the stellar populations compared to cosmic environment, in agreement with previous studies \citep[e.g.,][]{Erfanianfar2016,Yang2018b, Ni2021}.

\section{Discussion}
\label{sec_discussion}

Previous galaxy studies that have examined the role of the cosmic environment and morphology on the star-formation have found controversial results. For instance, \cite{Erfanianfar2016}, used X-ray-selected galaxy groups and found little dependence of the star-formation on environment at $\rm 0.5<z<1.1$, for main-sequence (MS) galaxies. Their findings suggest that stellar mass, morphology and environment act together in moving the star formation to lower levels. However, environmental effects are more prominent at lower redshifts. \cite{Darvish2016} used star-forming and quiescent galaxies in the COSMOS field at $\rm z\leq 3$. They found that star-forming galaxies have similar SFR and sSFR in different environments, regardless of their redshift and M$_*$. The fraction of Q systems depends on the environment at $\rm z\leq 1$ and on M$_*$ out to $\rm z\sim 3$. Moreover, galaxies at $\rm z\leq 1$ become Q faster in denser environments and the overall environmental quenching efficiency increases with cosmic time. They also concluded that environmental and mass quenching processes depend on each other and they become prominent at different cosmic time. Specifically, environmental quenching is only relevant at $\rm z\leq 1$, whereas mass quenching is the dominant mechanism at $\rm z\geq 1$. \cite{Leslie2020} used galaxies in the COSMOS field and found no variations in the main-sequence in different environments. Albeit, they found that BD galaxies have lower SFR than disk-dominated galaxies, at fixed M$_*$ at $\rm z<1.2$. \cite{Delgado2022} studied galaxy members in 80 groups at $\rm z\leq 0.8$ and found, among others, that the fraction of red and quiescent galaxies is always higher in groups than in the field. More recently, \cite{Cooke2023}, used star-forming galaxies in the COSMOS field and found no environmental and morphological dependence of the shape of the MS, up to redshift of 3.5. \cite{PerezMillan2023} performed a detailed study on the dependence of the observed galaxy properties on M$_*$, morphology and environment and found that at fixed M$_*$ and morphology, clusters have higher quenching efficiencies. They conclude that the cluster environment affects the ability to form stars, independent of morphological type. \cite{Sobral2022} used the LEGA-C catalogue and the local overdensities estimated by \cite{Darvish2015, Darvish2017} and found that the fraction of Q galaxies hinges on M$_*$ and on the environmental density. Their findings also showed that the D$_n$4000 and H$_\delta$ indices of galaxies depends on M$_*$ and on the cosmic environment, in particular, in the case of Q galaxies, whereas for SF systems the two spectral indices depend, mainly, on M$_*$ and do not show a significance dependence on environment. They concluded that Q galaxies in higher density fields are older, formed and/or quenched earlier.

Our analysis shows that galaxies with $\rm log\,[M_*(M_\odot)]>10.6$, at $\rm z\sim 1$, have different stellar populations in different cosmic environments  (Table \ref{table_sfh_numbers}) and these differences are, mainly driven by Q (or BD) systems (Table \ref{table_sfh_numbers_Q} and \ref{table_morphology}), in agreement with, for instance, \cite{Sobral2022}. Our results, also, show that galaxy morphology is affected by the environment, with the fraction of BD systems to increase with the density field \citep[see also e.g.,][]{Dressler1980,Fasano2015}. Recently, \cite{Hasan2023} analyzed outputs from the Illustris-TNG-100 magnetohydrodynamical cosmological simulations \citep{Pillepich2018, Nelson2019}. Their analysis showed that the influence of the cosmic environment on the star formation activity starts at $\rm z\sim 2$ and that most of the star formation-environment relationships can be explain in terms of the evolution of the median gas fraction. Specifically, the quenching of star formation in dense environments is due to rapid gas stripping or in the case of intermediate density fields due to gradual gas starvation. 

%We also find a stronger dependence of the SFH on morphology compared to environment, in agreement with previous studies \citep[e.g.,][]{Erfanianfar2016,Yang2018b, Ni2021}.

The main aspect of this work is the role of the SMBH activity in the star formation and stellar population of galaxies in different environments (and morphologies). Our results showed that galaxies that host low luminosity AGN ($\rm log\,[L_{X,2-10keV}(erg\,s^{-1})]<43$) have, on average, consistent stellar populations independent of the cosmic environment. Their SFR is also lower compared to the SFR of non-AGN galaxies (SFR$_{norm} \sim 1$), regardless of the field density. Galaxies that host moderate L$_X$ AGN ($\rm 43<log\,[L_{X,2-10keV}(erg\,s^{-1})]<44$) have stellar populations that tend to be younger in denser fields compared to the field. Their SFR is also similar, if not higher, compared to the SFR of non-AGN galaxies (SFR$_{norm} \geq 1$). This could imply that the AGN feedback in dense environments counter-acts the removal of the gas and prevents the quenching of the star formation \citep[positive feedback; e.g.,][]{Zinn2013, Santoro2016, Meenakshi2022}. The higher the AGN activity (L$_X$), the most efficient the AGN feedback. However, an alternative interpretation of our results could be that a common mechanism \citep[for instance mergers, e.g.,][]{DiMatteo2005,Hopkins2008a} feeds both the star formation and the SMBH and triggers the SF and the AGN activity \citep[e.g.,][]{Bower2006, Fanidakis2012}. In this scenario, the more gas funnelled to the galaxy due to the triggering mechanism, the higher the increase of the SFR and the AGN power (L$_X$). This picture is also consistent with the different morphological types that prevail in different environments between AGN and non-AGN systems. Specifically, the rarity of BD galaxies in systems that host AGN in dense fields, could be an indication of galaxy interactions that trigger both phenomena. Finally, the lower SFR of galaxies that host AGN compared to non-AGN systems (SFR$_{norm}<1$), in low dense fields, could be explained by low availability of gas in these systems. For instance, \cite{Zubovas2013} showed that in galaxies that are in a gas-poor phase AGN feedback may quench the star formation.

\section{Conclusions}
\label{sec_conclusions}

We used 551 X-ray AGN and 16917 non-AGN galaxies, at $\rm 0.3<z<1.2$, in the COSMOS field to compare their SFR (SFR$_{norm}$) as a function of L$_X$, in different cosmic environments. We restrict the M$_*$ of both population to the range of $\rm 10.5<log\,[M_*(M_\odot)]<11.5$. The classification of the sources in three environments (field, filaments, clusters) is available from the catalogue of \cite{Yang2018b}. The categorization is based on their field density using the method described in \cite{Darvish2015}. The (host) galaxy properties were calculated via SED fitting, using the CIGALE code and the measurements have been presented in \cite{Mountrichas2022a}. We then, cross-matched our datasets with the LEGA-C catalogue that allowed us to study the SFH of the two populations, for different field densities. This information is available for 52 AGN and 1172 non AGN galaxies. Finally, we compared the role of environment vs. morphology, using the morphological information in the catalogue of \cite{Ni2021}, using 37 AGN and 1086 non-AGN galaxies that are common between the datasets. Our main results can be summarized as follows:

\begin{itemize}

\item[$\bullet$] The SFR of galaxies that host AGN is lower compared to the SFR of non-AGN systems (SFR$_{norm}$<1), for isolated sources (field), at all L$_X$ spanned by our dataset ($\rm 42<log\,[L_{X,2-10keV}(erg\,s^{-1})]<44)$. However, in denser environments an increase of SFR$_{norm}$ (SFR$_{norm}\sim 1$) is observed at intermediate L$_X$ ($\rm log\,[L_{X,2-10keV}(erg\,s^{-1})]>43$).

\item[$\bullet$] SFR$_{norm}$ appears higher in fields of higher density compared to lower density fields.
%, at a significance level of $\sim 2\sigma$.

\item[$\bullet$] Non-AGN galaxies tend to have older stellar populations (higher D$_n$4000 values) and are less likely to have undergone a recent star-formation burst (lower H$_\delta$ values) in denser environments compared to isolated sources (field). On the contrary, low L$_X$ AGN have consistent stellar populations (similar D$_n$4000 and H$_\delta$ values), in all cosmic environments, while moderate L$_X$ AGN tend to have younger stars and are more likely to have undergone a recent burst, in high density fields.

\item[$\bullet$] The differences of the stellar populations for different cosmic environments are, mainly, driven by quiescent systems, both in the case of AGN and non-AGN galaxies.

\item[$\bullet$] Although the same fraction of AGN and non-AGN systems is associated with different cosmic environments and the same fraction of BD systems is found both for AGN and non-AGN galaxies, the morphology as a function of the environment is different for the two populations. Specifically, in the case of non-AGN galaxies, the percentage of BD systems increases with the field density, whereas in the case of AGN, BD systems become scarce in denser environments.

%\item[$\bullet$] The link between morphology and galaxy phase is weaker for AGN compared to non-AGN galaxies.

%\item[$\bullet$] Morphology appears to be a more determinative factor compared to the cosmic environment regarding the stellar populations of AGN and, mainly of, non-AGN galaxies.

\end{itemize}

In this work, we studied the SFR and stellar populations of AGN and non-AGN galaxies in different cosmic environments (and morphologies). Larger samples that include sources observed in wider fields, such as the eFEDS field, would be extremely useful to confirm our findings and expand our investigations to larger M$_*$, L$_X$ and redshift baselines. Previous studies have showcased the importance of M$_*$ when comparing the star formation and stellar populations of AGN and non-AGN systems. Therefore, larger samples that span, in sufficient numbers, lower ($\rm log\,[M_*(M_\odot)]<10.5$) and higher ($\rm log\,[M_*(M_\odot)]>11.5$) M$_*$ are important to examine if our results change at different M$_*$ regimes. Datasets that cover a wider redshift baseline, will also allow to examine a possible dependence with cosmic time. Finally, the X-ray AGN used in this study, probe low-to-moderate L$_X$. It would be particularly interesting to check if and how the results change in the case of the most powerful AGN ($\rm log\,[L_{X,2-10keV}(erg\,s^{-1})]>44$).

\begin{acknowledgements}
This project has received funding from the European Union's Horizon 2020 research and innovation program under grant agreement no. 101004168, the XMM2ATHENA project.
The project has received funding from Excellence Initiative of Aix-Marseille University - AMIDEX, a French 'Investissements d'Avenir' programme.
This work was partially funded by the ANID BASAL project FB210003. MB acknowledges support from FONDECYT regular grant 1211000.
This research has made use of TOPCAT version 4.8 \citep{Taylor2005}.

\end{acknowledgements}

\bibliography{mybib}
\bibliographystyle{aa}

\end{document}